# Vehicle Communication using Secrecy Capacity

Na-Young Ahn[1], Donghoon Lee[2] and Seong-Jun Oh[2]

1 Graduate School for Information Security, Korea University, Seoul, Korea
2 Professor at CIST and Graduate School for Information Security, Korea University

```
humble@korea.ac.kr
```

**Abstract.** We address secure vehicle communication using secrecy capacity. In particular, we research the relationship between secrecy capacity and various types of parameters that determine secrecy capacity in the vehicular wireless network. For example, we examine the relationship between vehicle speed and secrecy capacity, the relationship between the response time and secrecy capacity of an autonomous vehicle, and the relationship between transmission power and secrecy capacity. In particular, the autonomous vehicle has set the system modeling on the assumption that the speed of the vehicle is related to the safety distance. We propose new vehicle communication to maintain a certain level of secrecy capacity according to various parameters. As a result, we can expect safer communication security of autonomous vehicles in 5G communications.

## 1    INTRODUCTION

Intelligent transportation systems (ITSs) can improve road safety and traffic efficiency for any smart city [1]. Vehicular wireless network security is of vital importance for the deployment of ITSs in practice [2], [3]. Previous works mainly focus on the use of co-operative relaying to enhance transmission reliability, the improvement of end-to-end throughput, and the extension of the service range of vehicular networks. However, the openness of the vehicular channels makes transmission data available to illegal users as well as the eavesdroppers. Therefore, guaranteeing the secrecy of the data transmission is also of vital importance. IEEE 1609.2 specifies the formats for secure messages and the corresponding crypto procedure in the vehicular wireless networks. However, management of the secret keys often requires a trusted third party authorization as well as complex network architectures and protocols that are difficult to satisfy in vehicular wireless networks [4]. Ao Lei introduced a Blockchain-based dynamic key management for heterogeneous ITSs [5]. Instead of the cryptography based security, physical layer security (PLS) has recently been introduced by [6], [7], [8].

We have more deeply studied V2V (vehicle-to-vehicle) communication using physical layer security based on secrecy capacity. This paper is described as follows. First, the relationship between the safety distance and the speed of the vehicle is studied. Next, a system for calculating secrecy capacity according to the vehicle speed is modeled. After that, we have looked at the relationship of secrecy capacity according to various parameters. Next, the concept of vehicle communication using secrecy capacity is disclosed. Finally, in addition to physical layer security according to secrecy capacity in vehicle communication, we have also studied how to increase physical layer security.

## 2    SAFETY DISTANCE BASED ON VELOCITY

In self-driving automation, Adaptive Cruise Control (ACC) is a very important technique. ACC is configured to regulate vehicle speed and distance between at least two vehicles, and systems may have to meet conditions of collision avoidance in various situations. For example, there are three scenarios: Stop and Go, Emergency braking, and Cut-in [9]. All scenarios include the braking distance between a preceding vehicle and a following vehicle. Generally, the braking distance refers to a distance that a vehicle will run from an initial braking point to a complete stop. The braking distance includes a safety distance between a host vehicle and a target vehicle [10]. That is, the safety distance is at least as long as the braking distance.

The autonomous vehicle may perform the braking process shown referring to FIG. 1.

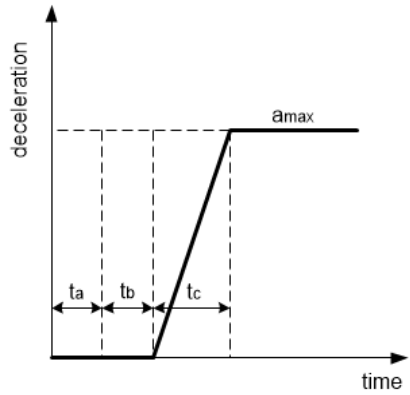

**Figure 1. Braking Process in the host vehicle**

In FIG.1, $a_{max}$ is the maximum deceleration in the braking process. The braking process comprises the response period $t_a$, the braking clearance period $t_b$ and the breaking force applying period $t_c$. In this time, theoretical braking distance $d_L$ is calculated as follows [11]:

$$d_L = v_0 \left(t_a + t_b + \frac{t_c}{2}\right) + \frac{v_0^2}{2a_{max}}$$

where $v_0$ is the initial velocity of the host vehicle.

In FIG.2, the braking safety distance is shown to prevent a collision between the host vehicle A and the following vehicle B.

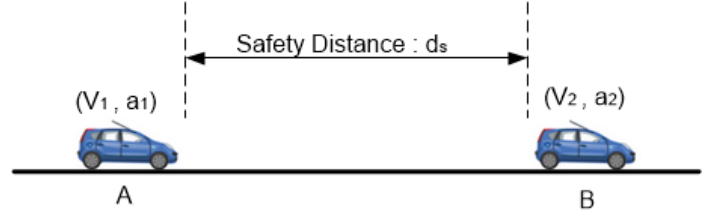

**Figure 2. Vehicle Safety Distance for preventing an end collision**

As shown in FIG. 2, when the host vehicle A detects that two vehicles (A, B) may collide, it will automatically operate a safe guard in the intelligent cruise control system. Now assume that the host vehicle A is at initial velocity $V_1$, acceleration $a_1$; the target vehicle B is at initial velocity $V_2$, acceleration $a_2$; the safety distance is $d_s$. The safety distance $d_s$ is calculated as follows [12]:

$$d_s = v_1 \tau + \frac{V_1^2}{2a_1} - \frac{V_2^2}{2a_2}$$

where τ is a constant parameter of the intelligent cruise control system.

Our main concern is secure V2V communication. Therefore, we assumed that the host vehicle A and the target vehicle B based on ACC systems have constant motion and the same speed. Because the respective ACC systems of the host vehicle A and the target vehicle B control to maintain the spacing distance between the first vehicle A's front bumper and the second vehicle B's rear bumper, $V_1 = V_2$, and $a_1 = a_2$. As a result, the safety distance based on the host vehicle A's velocity is represented below:

$$d_s = v_1 \tau .$$

Consequently, we confirmed that the safety distance $d_s$ is proportional to the initial velocity $V_1$ of the host vehicle A in V2V communication. That is, the speed of the autonomous vehicle is proportional to the safety distance.

## 3 SECRECY CAPACITY FOR V2V COMMUNICATION

In information theory, channel (or Shannon) capacity is known as the maximum amount of information that can be transmitted through a wireless channel. In general, channel capacity is given as:

$$C = W \log(1 + \text{SNR})$$

where W is the channel bandwidth and SNR is the signal-to-noise ratio. Secrecy capacity means that the channel capacity of a legitimate channel subtracts the channel capacity of a wiretap channel. That is, secrecy capacity is the maximum data rate achievable between the legitimate TX-RX pair, subject to the constraints on information attainable by the unauthorized receiver [13]. In the Gaussian wiretap channel, secrecy capacity $C_s$ is given by

$$C_s = \frac{1}{2} \log \left(1 + \frac{P}{N_m}\right) - \frac{1}{2} \log \left(1 + \frac{P}{N_w}\right)$$

where P is a transmitter's power, $N_m$ is a receiver's noise, $N_w$ is an eavesdropper's noise.

### 3.1 SYSTEM MODEL BASED ON VELOCITY

We expect that the distance between the host vehicle A and the target vehicle B is longer than the safety distance $D_s$ in the autonomous vehicle system. We assumed that the eavesdropper E is far away relative to the host vehicle A and the target vehicle B. That is, the distance r between the host vehicle A and the eavesdropper E is similar to the distance r' between the target vehicle B and the eavesdropper E. The distance D between the host vehicle A and the target vehicle B is rθ, where θ is an angle that is formed by a first line AE and a second line AE, as shown in FIG. 3.

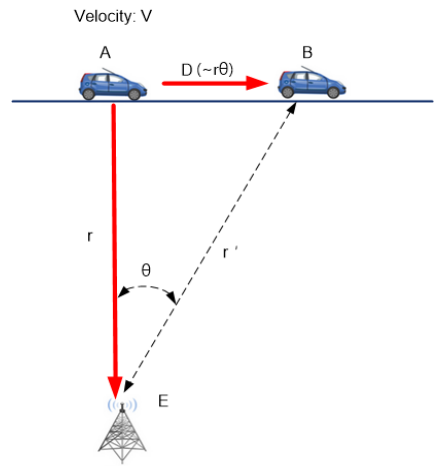

**Figure 3. System Model under faraway eavesdropper**

We consider a V2V communication scenario with the eavesdropper E. Secrecy capacity $C_s$ in the fading scenario is given by [14], [15], [16], [17]

$$C_s = \log_2\left(1 + \frac{P|h_{AB}|^2}{N_0}\right) - \log_2\left(1 + \frac{P|h_{AE}|^2}{N_0}\right)$$

where P is the transmission power of the host vehicle A, $h_{AB}$ is a fading channel coefficient between the host vehicle A and the target vehicle B, $h_{AE}$ is a fading channel coefficient between the host vehicle A and the eavesdropper E, and $N_0$ is the variance of additive white Gaussian noise (AWGN).

Generally, there are three fading models: Rayleigh, Rician, and Nakagami. The transmitted signal power may reduce according to the distance as $d^{-\alpha}$, where $\alpha$ is a path loss exponent [18]. When the path loss distance is the distance D (which is equal to $r\theta$) between the host vehicle A and the target vehicle B, the fading channel coefficient $h_{AB}$ is given by

$$h_{AB} = \left|\frac{1}{(r\theta)^\alpha}\right|$$

.

Also, when path loss distance is the distance r between the host vehicle A and the eavesdropper E, the fading channel cofficient $h_{AE}$ is given by

$$h_{AE} = \left|\frac{1}{r^\alpha}\right|$$

.

Accordingly, in this system model secrecy capacity $C_s$ is given by:

$$C_s = \log_2\left(1 + \frac{P}{N_0(r\theta)^{2\alpha}}\right) - \log_2\left(1 + \frac{P}{N_0 r^{2\alpha}}\right)$$

.

We found that the velocity of the host vehicle is associated to the safety distance $d_s$ between the legitimated terminals as: $d_s = v_1 \tau$. From a velocity point of view, the distance D between the host vehicle A and the target vehicle B is given by:

$$D = r\theta = v\tau$$

where v is the current velocity of host vehicle A and $\tau$ is a constant parameter of the ACC system.

As a result, in this system model, secrecy capacity $C_s$ is given by:

$$C_s = \log_2\left(1 + \frac{P}{N_0(v\tau)^{2\alpha}}\right) - \log_2\left(1 + \frac{P}{N_0 r^{2\alpha}}\right)$$

.

For the simplicity of analysis, we assumed that the distance r is fixed. Then secrecy capacity $C_s$ is a function expressed by only two variable parameters v and $\alpha$. FIG. 4 shows secrecy capacity $C_s$ according to vehicle velocity v.

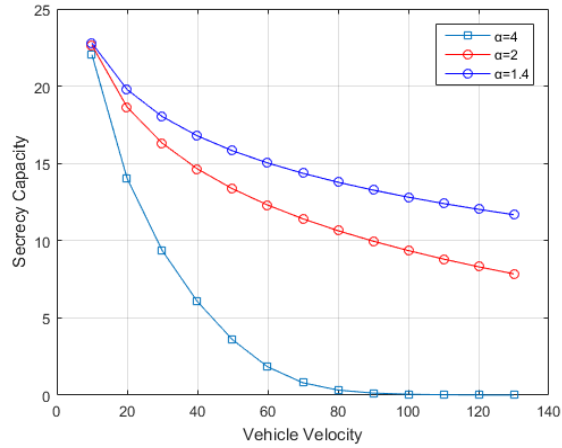

**Figure 4. Secrecy Capacity Variances according to vehicle speed having various path loss coefficients ($\alpha$=4, $\alpha$=2, and $\alpha$=1.4) with r=1000 m, $P/N_0$=70dB, and $\tau$= 200ms**

As shown in FIG. 4, we found that secrecy capacity $C_s$ is greatly affected by the path loss coefficient $\alpha$ and the vehicle velocity v. First, when the vehicle velocity v increases, the secrecy capacity $C_s$ decreases regardless of the path loss coefficient $\alpha$. Second, the greater the path loss coefficient $\alpha$, the greater the secrecy capacity $C_s$.

We also expect that an increase of velocity in the host vehicle A leads to a decrease of secrecy capacity $C_s$. For convenience of explanation, we assume that autonomous vehicles are traveling on highways. In simulation, FIG. 5 shows the secrecy capacity $C_s$ according to different velocities: 80 Km/h, 100 Km/h, and 120 Km/h. In this simulation, we assumed that the distance r between A and E is 1000m, and the fading channel model is the Rayleigh fading channel model. As shown in FIG. 4, the secrecy capacity is highest at 80 Km/h and lowest at 120 Km/h. We confirmed that a high speed of the vehicle may reduce the secrecy capacity. On considering security environments, we aim to optimize the transmission power P or the transmission rate according to the vehicle speed.

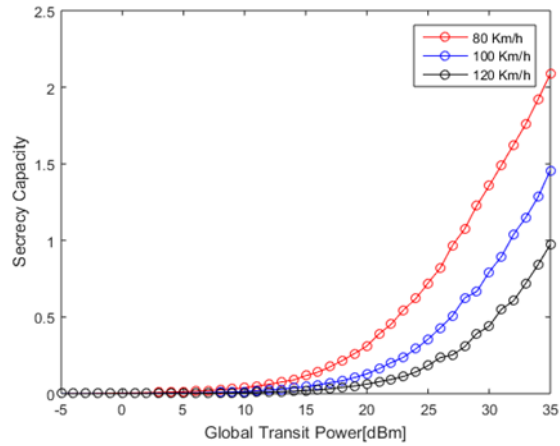

**Figure 5. Secrecy Capacity Variances at Vehicle speed (80 Km/h, 100 Km/h, and 120 Km/h) with Path loss coefficient: α=3.5, r=1000m, and θ ≒ 0.1**

As shown in FIG. 6, we can find a change in secrecy capacity with transmission power ratio $P/N_0$. It can be seen that the secrecy capacity is proportional to the magnitude of the transmission power ratio $P/N_0$. That is, the greater transmission power ratio $P/N_0$, the greater secrecy capacity $C_s$, referring to FIG. 6.

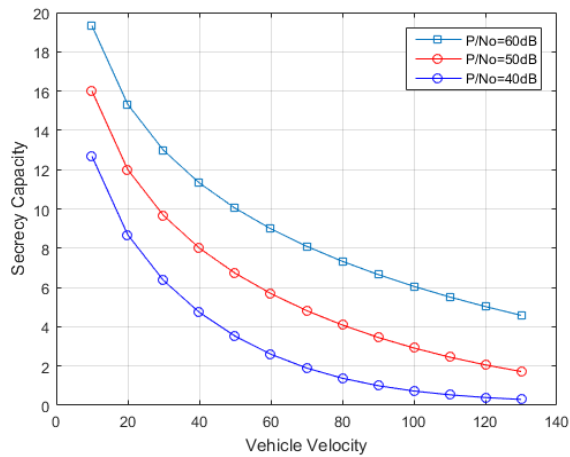

**Figure 6. Secrecy Capacity Variances according to Vehicle speed at Transmission Power ($P/N_0$=40dB, $P/N_0$=50dB, and $P/N_0$=60dB) with α=1.4, r=1000 m, and τ= 400ms**

In addition, we examined the secrecy capacity of the system according to the reaction speed of the ACC system. As shown in FIG. 7, it is confirmed that the faster the response speed of the system, the greater the secrecy capacity.

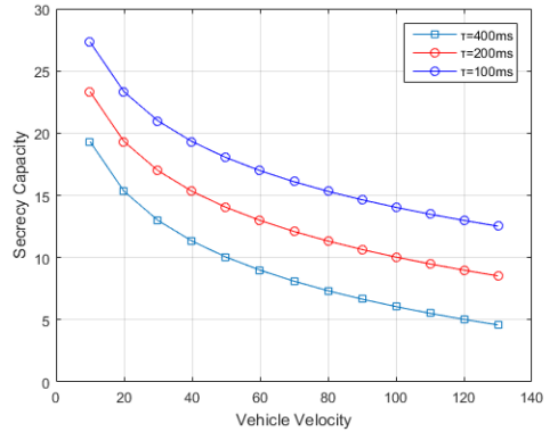

**Figure 7. Secrecy Capacity Variances according to Vehicle speed at a response speed (τ= 100ms, τ= 200ms, τ= 400ms) with α=1.4, and r=1000 m**

Accordingly, we have found that it is important to select appropriate vehicle velocities and transmission power ratios $P/N_0$ to keep a constant secrecy capacity. This means that we can define the secrecy capacity as a criterion to maintain a certain level of security in vehicular communications.

As described above, secrecy capacity can be determined by the vehicle speed, the magnitude of the transmission power, and the system parameters. In summary, the schematic orientation of secrecy capacity can be determined by the following table.

**Table 1 shows the relationship between various parameters and secrecy capacity.**

| Vehicle Speed | Up | Down | Secrecy Capacity |
|---|---|---|---|
|  | Down | Up |  |
| Transmission Power | Up | Up |  |
|  | Down | Down |  |
| System Parameter (Response Time) | Up | Down |  |
|  | Down | Up |  |

## 3.2 SYSTEM MODEL BASED ON RELAY

As described above, secrecy capacity decreases as the vehicle speed increases, but it can be compensated by cooperative relay communication. In V2V communication, secrecy capacity can be improved by introducing one relay R between the host vehicle A and the target vehicle B. In general, secrecy capacity of cooperative relay communication is higher than that of direct communication without a relay [19]. For simplicity of the analysis of secrecy capacity, we assume that the system model comprises one relay R between the host vehicle A and the target vehicle B, as shown in FIG. 8.

Channel capacity of the legitimated channel is expressed as:

$$C_1(A, B) = W \log_2 \left(1 + \left(\frac{P_A h_{AB}}{P_R h_{RB} + \sigma_B^2}\right)\right)$$

where $P_A$ and $P_R$ are the transmission powers of the host vehicle A and the relay R, respectively, $h_{AB}$ is the channel gain between the host vehicle A and the target vehicle B, $h_{RB}$ is the channel gain between the relay R and the target vehicle B, $\sigma_B^2$ is an additive white Gaussian noise at the target vehicle B, and W is a bandwidth.

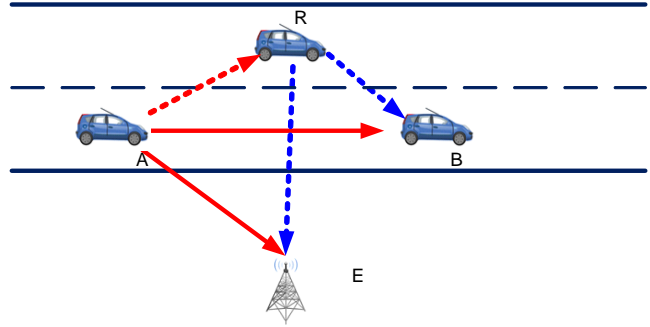

**Figure 8. System Model with a Relay between A and B**

Now, the channel capacity of the wiretap channel is given by:

$$C_2(A, E) = W \log_2 \left(1 + \left(\frac{P_A h_{AE}}{P_R h_{RE} + \sigma_E^2}\right)\right)$$

where $h_{AE}$ is the channel gain between vehicle A and eavesdropper E, and $\sigma_E^2$ is the additive white Gaussian noise at vehicle B. Then, secrecy capacity with cooperative relay communication is denoted by

$$C_R = W \left[\log_2 \left(1 + \left(\frac{P_A h_{AB}}{P_R h_{RB} + \sigma_B^2}\right)\right) - \log_2 \left(1 + \left(\frac{P_A h_{AE}}{P_R h_{RE} + \sigma_E^2}\right)\right)\right]$$

.

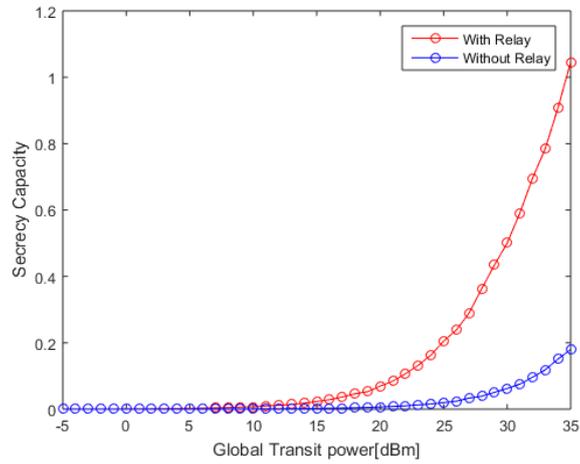

**Figure 9. Secrecy Capacity Variance according to Relay Existence**

FIG. 9 shows secrecy capacity with and without the relay R. Referring to FIG. 9, we know that the relay R helps to improve secrecy capacity overall. We confirmed that V2V communication using the relay may enhance secrecy capacity. The relationship between the existence of the relay and secrecy capacity can be summarized as the following table.

**Table 2 shows the relationship between the relay mode and secrecy capacity.**

| Relay | Off | Down | Secrecy Capacity |
|---|---|---|---|
|  | On | Up |  |

## 3.3 ERGODIC SECRECY CAPACITY

Secrecy capacity does not consider the channel variation over time. So, we introduce the concept of ergodic secrecy capacity. Ergodic secrecy capacity is defined as the time average of the secrecy rate over the legitimated user and the eavesdropper [20], [21], [22]. For an ergodic fading channel in MIMO systems, the fading channel coefficients are independent and identically distributed. Thus, in the vehicular network, each of the host vehicle A, the target vehicle B, and the eavesdropper E may experience a different fading state for respective channel use. Assuming that all terminals have perfect CSI (channel state information) about the current fading state (CSI), ergodic secrecy capacity is denoted by:

$$C_S = \max_{E_A[\gamma] \leq P} E_A \, log_2\left(1 + \frac{\gamma |h_{AB}|^2}{\sigma_B^2}\right) - log_2\left(1 + \frac{\gamma |h_{AE}|^2}{\sigma_E^2}\right)$$

with $\gamma$ the power allocation, and

$$A = \left\{h_{AB}, \; h_{AE} : \frac{|h_{AB}|^2}{\sigma_B^2} > \frac{|h_{AE}|^2}{\sigma_E^2}\right\}$$

[23].

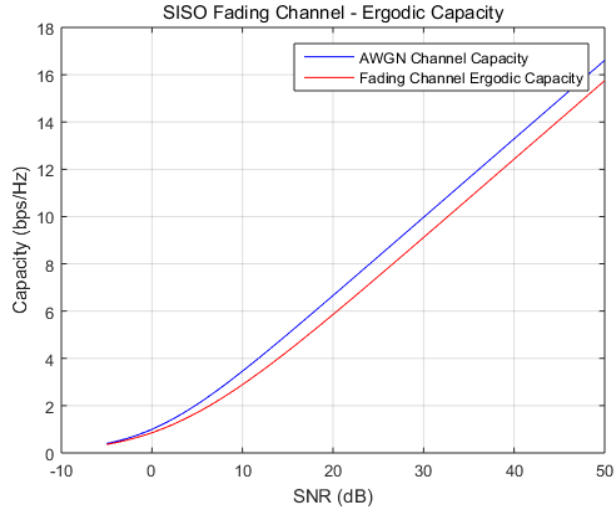

**Figure 10. Ergodic Secrecy Capacity vs. AWGN Capacity, Noise power is assumed to be unity.**

.

We compare AWGN (additive white Gaussian noise) channel capacity with the fading ergodic channel capacity. As shown to FIG. 10, the AWGN channel capacity is higher than the ergodic capacity.

## 4  PROPOSED V2V COMMUNICATION USING SECRECY CAPACITY

Recent studies show an increased interest in calculating the vehicular communication capacity in 5G mobile networks [24], [25], [26], [27]. As described above, we confirm the relationship between vehicle speed and secrecy capacity, as well as the relationship between relay existence and secrecy capacity. We propose new V2V communication that performs with cooperative relay communication according to a value of secrecy capacity, referring to FIG. 11.

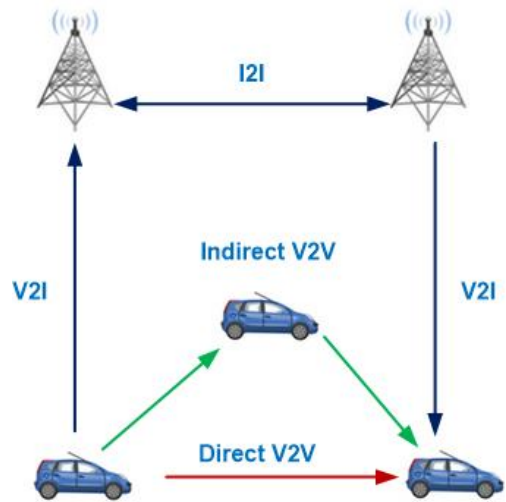

**Figure 11. Proposed Vehicle Communications**

Our proposed V2V communication process is as follows:
   1. A host vehicle receives CSI (channel state information) from a target vehicle.
   2. The host vehicle calculates secrecy capacity for connecting the target vehicle. Here, secrecy capacity is calculated by a predetermined scheme.
   3. The host vehicle determines whether secrecy capacity exceeds the predetermined value. Here, the predetermined value may be changed according to the speed range of the vehicle.
   4. If secrecy capacity exceeds the predetermined value, the host vehicle may initiate direct communication with the target vehicle.
   5-1. In one case, if the secrecy capacity does not exceed the predetermined value, the host vehicle may initiate indirect communication with the target vehicle through cooperative relay communication. Then, a relay selection of the vehicular network is performed by an optimization algorithm that considers power consumption, secrecy enhancement, etc. For example, the optimization algorithm can control the transmission power of the host vehicle to increase security in V2V communication. The above V2V communication induces the enhancement of physical layer security due to cooperative relay communication based on secrecy capacity.

5-2. In another case, if the secrecy capacity does not exceed the predetermined value, the host vehicle may increase the transmission power of the host vehicle by the predetermined amount. Consequently, vehicle communication will be initiated.

Meanwhile, the host vehicle can select V2I (Vehicular to Infrastructure) communication, such as 5G mobile networks, according to the result of the secrecy capacity.

In addition, we will define new CSI (channel state information) for 5G vehicle communication. The main content that we propose is included in the SNR value in the channel state information. Trying to perform initial communication, it is not difficult to calculate the SNR value at the receiver if the receiver and receiver know the power to transmit because they know the received power. All receivers will be able to calculate SNR values and include them in the channel state information. For this proposal, we first need to have enough discussion in the relevant standards associations.

## 5 PHYSICAL LAYER SECURITY ENHANCEMENT

In short, V2V communication satisfies physical layer security based on the secrecy capacity. Additionally, we can enforce the security using a physical layer encryption based on compressive sensing. Compressive sensing encryption can be an attractive solution as it can provide reasonably secure transmissions with simple low-complexity ciphers in the physical layer [28-37]. Generally, compressive sensing schemes can sample signals below the Nyquist rate [38].

According to V2V communication between the host vehicle A and the target vehicle B, a compressive sensing encryption can be optional, referring to FIG. 12. The host vehicle A can determine whether or not using the compressive sensing encryption is desirable based on the value of secrecy capacity.

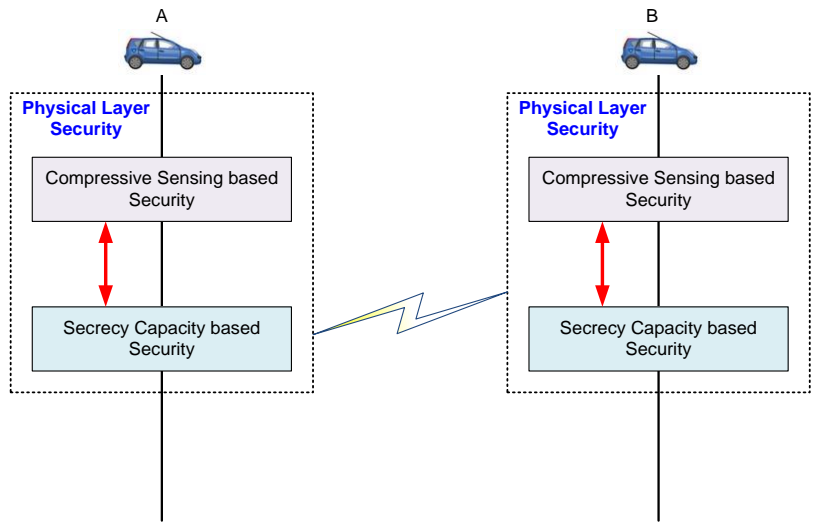

Figure 12.   Advanced Physical Layer Security based on Compressive Sensing Encryption

# 6 CONCLUSION

We studied the relationship between secrecy capacity and the speed of a vehicle in the vehicular network. As a result, we confirmed that a high speed of the vehicle may reduce secrecy capacity. On considering security environments, we need to optimize a transmission power or a transmission rate according to the vehicle speed. We also confirmed that V2V communication using the relay may enhance secrecy capacity. Accordingly, the vehicle user needs to perform V2V communication through the optimized selection relay algorithm, which selects the relay to increase secrecy capacity based on the vehicle's velocity. In addition, we introduced a physical layer security using compressive sensing encryption. In the future, we may lead to improvements, such as more efficient design and faster, more secure communication in the vehicular network through physical layer security.